# Manifestation of the magnetic moments of Cooper pairs in low-temperature properties of superconducting thin-film rings

A.I. Agafonov

NRC "Kurchatov Institute", Kurchatov sq. 1, Moscow 123182, Russia

E-mail: aai@isssph.kiae.ru

**Abstract**. Experimental determination of the magnetic moment of the Cooper pairs can shed light on the pairing symmetry in cuprates. We argue that the intrinsic magnetic moments of the Cooper pairs can be detected in experiments with superconducting thin-film rings. At sufficiently low temperatures the magnetic field generated by the supercurrent, can cause the ordering of these magnetic moments that produces changes in the supercurrent and magnetic induction distributions, the heat capacity and magnetic moment of the ring. It is shown that the intrinsic magnetic moment of the Cooper pairs can be extracted from low-temperature behaviours of these measurable quantities of the current-carrying rings made of the cuprate superconductors.

## 1. Introduction

The pairing symmetry in the cuprate superconductors is still controversial topic [1,2]. From the theoretical point of view, the d-wave pairing is mostly argued [3], though another spin-singlet channels, such as the s-wave pairing and admixed $d_{x^2-y^2}+s$ pair state are also discussed [1,2,4,5]. As noted in [6], the central symmetry of the $CuO_2$ planes can be broken down because of asymmetric surroundings of these planes in the parent compounds, as occurs in $YBa_2Cu_3O_{7-\delta}$. Also, doping-induced disorder can lead to the absence of this symmetry in the doped cuprates [7]. As a result, the mixed singlet-triplet state in the cuprates may be, in principle, possible [7,8,9].

High hopes to solve the problem of pairing symmetry are entrusted to the bulk- and phase-sensitive experimental techniques based on the macroscopic quantum coherent effects in superconductors. Though the $d_{x^2-y^2}$ pair state have been actively sought in numerous studies [1,2,10,11], a variety of experimental tests yielded conflicting results [12-15]. Therefore, another way for determining this symmetry is seemed to be urgent.

Here we focus on the well-known aspect that the pairing symmetry in superconductors and the intrinsic magnetic moment of the Cooper pairs are interrelated. So, for the $s$-wave pairing predicted by the Bardeen-Cooper-Schrieffer theory, the relative orbital angular momentum of the electron pairs is $l = 0$, and their orbital and spin magnetic moments are equal to zero. The most common opinion is that the cuprates inherent singlet $d$-wave pairing. This means that the orbital angular momentum of the electron pairs $l = 2$, and, respectively, the pairs have the orbital magnetic moment. For the mixed singlet-triplet state, besides the intrinsic orbital magnetic moment, the Cooper pairs have the spin magnetic moment as well.

Thus, the experimental determination of the magnetic moment of the Cooper pairs in superconductors and, in particular, the cuprates can shed light on the pairing symmetry. In addition, it is important to establish whether this symmetry is the same in different regions of the phase diagram of doped cuprates. For this purpose it is necessary to find experimental conditions in which the magnetic moments of the Cooper pairs will manifest themselves in observables.

In this paper we show that the magnetic moments of the Cooper pairs can be determined in experiments with thin-film rings made of the cuprate superconductors. The magnetic field generated by the persistent supercurrent in the ring, penetrates into the superconductor, if the ring thickness is less than the London penetration depth. In the case of sufficiently low temperatures this magnetic field causes the ordering of the magnetic moments of the Cooper pairs that leads to the local magnetization of the superconductor. This paramagnetic response produces changes in low-temperature properties of the ring such as the supercurrent and magnetic field distributions, the heat capacity and magnetic moment of the superconducting ring. The main result of our study is that from experimental measurements of the low temperature dependences of these observables one can determine the magnetic moment of the Cooper pairs.

**2. System**

Consider a flat thin-film ring made of the cuprate superconductor, with the inner radius $a$, the outer radius $b$ and the ring thickness $d$. In the cylindrical coordinates used below, the $z$-axis coincides with the crystallographic $c$-axis, and the ring occupies the region $a \leq \rho \leq b$ and $-d/2 \leq z \leq d/2$. In this geometry the currents flow in the $CuO_2$ planes. It is assumed that the magnetic field generated by the supercurrent in the ring, is less than the first critical field for the superconductor. This imposes a limit on the number of fluxoids trapped into the ring. In our case, the current and magnetic field distributions in the ring are determined by the penetration depth $\lambda_{ab}$.

Calculations are carried out for the rings with $a \gg \lambda_{ab}$, $b - a \gg \lambda_{ab}$ and $d < \lambda_{ab}$. The latter allows to neglect the $z$-dependence of the current density and magnetic field into the superconducting film. Because of the circular symmetry, the current density and vector-potential have only a $\varphi$-component, $\mathbf{j}(\rho) = j(\rho)\mathbf{i}_\varphi$ and $\mathbf{A}(\rho) = A(\rho)\mathbf{i}_\varphi$, where $\mathbf{i}_\varphi$ is the azimuthal unit vector. The magnetic induction in the film and ring hole has only a $z$-component, $\mathbf{B}(\mathbf{r}) = B(\rho)\mathbf{i}_z$.

In the case of the $d$-wave symmetry, the Cooper pair has the intrinsic orbital magnetic moment $\boldsymbol{\mu}_i$. Its $z$-projection is

$$\mu_z = \mu_C m_L, \tag{1}$$

where $\mu_C$ is the order of the Bohr magneton and $m_L = 0, \pm 1, \pm 2$.

For the isotropic pairing the Cooper pair has no intrinsic magnetic moment.

## 3. Closed system of equations

As is well known, cuprates exhibit coherent lengths in the nanometer range. The magnetic field in the ring does not change on such lengths. Hence, for description of the Zeeman energy of the Cooper pairs and their interaction between each other we can use their center-of-mass radius-vectors. The density matrix can be presented as:

$$\rho = e^{-(W_Z + W_d)/kT} / Sp\left(e^{-(W_Z + W_d)/kT}\right), \tag{2}$$

where $T$ is the ring temperature, $W_Z$ is the Zeeman energy,

$$W_Z = -\sum_i \boldsymbol{\mu}_i \mathbf{B}(\mathbf{r}_i), \tag{3}$$

and $W_d$ is the energy of the magnetic dipole interaction of the Cooper pairs,

$$W_d = \frac{\mu_0}{2} \sum_{i \neq j} \left[ \frac{\boldsymbol{\mu}_i \boldsymbol{\mu}_j}{|\mathbf{r}_i - \mathbf{r}_j|^3} - \frac{3(\boldsymbol{\mu}_i(\mathbf{r}_i - \mathbf{r}_j))(\boldsymbol{\mu}_j(\mathbf{r}_i - \mathbf{r}_j))}{|\mathbf{r}_i - \mathbf{r}_j|^5} \right]. \tag{4}$$

Here $\mathbf{r}_i$ is the center-of-mass radius-vector of $i$-electron pair.

Using (2)-(4), methods were developed to calculate the average $A = Sp(\rho \hat{A})$ of an operator $A$ in form of a power series in $T^{-1}$ (see [16] and references therein). However, with respect to our problem, for not too low temperatures the energy (4) can be neglected as compared to (3). Evaluation of these temperatures is carried out in Section 4. Considering that in a real situation $T(<T_c) \gg \mu_C B(\rho = a)/k$, from (1), (2) and (3) we obtain the average magnetic moment of the Cooper pair:

$$\mu_z(\mathbf{r}_i) = Sp(\rho\hat{\mu}_z) = \frac{2\mu_C^2 B(\mathbf{r}_i)}{kT}. \tag{5}$$

This ordering of the magnetic moments of Cooper pairs leads to the magnetization vector $\mathbf{J}(\rho) = J_z(\rho)\mathbf{i}_z$ in the ring. Using (5), we obtain:

$$J_z = \frac{2\mu_C^2 n_C B(\rho)}{kT}. \tag{6}$$

where $n_C$ is the density of Cooper pairs.

From (6) we find the relationship between the magnetic induction and magnetic field strength:

$$\mathbf{B} = \mu_0 \frac{T}{T-T_0}\mathbf{H} \tag{7}$$

where $T > T_0$, and $T_0$ is the characteristic temperature, which depends on the intrinsic magnetic moment of the Cooper pairs,:

$$T_0 = \frac{2\mu_0\mu_C^2 n_C}{k}. \tag{8}$$

In the derivation of the characteristic temperature (8) we neglected the energy of the magnetic dipole interaction of the Cooper pairs, which is proportional to $(T_0/T)^2$, as shown in Section 4. At $T \approx T_0$, this energy is the same order of magnitude as the Zeeman energy for all the Cooper pairs. Therefore, further we assume that the ring temperature $T \gg T_0$.

Now let us estimate $T_0$. For $\mu_C = \mu_B$ and the density of the Cooper pairs $n_C = 10^{20} cm^{-3}$ we obtain $T_0 = 1.6 mK$. Of course, this characteristic temperature is too low compared with the superconducting transition temperature of the cuprates. Therefore there is a region of the ring temperature $T_0 \ll T \ll T_c$ in which all properties of the superconductor such as the penetration depth and density of the Cooper pairs, can be considered as constant.

From the Maxwell equation $rot\mathbf{H} = \mathbf{j}$ and (7), we obtain:

$$rot\mathbf{B} = \mu_0 \frac{T}{T-T_0}\mathbf{j}. \tag{9}$$

Introducing $\mathbf{B} = rot\mathbf{A}_j$, where $\mathbf{A}_j$ is the vector potential generated by the supercurrent, from (9) we have:

$$\mathbf{A}_j(\mathbf{r}) = \frac{\mu_0 T}{4\pi(T-T_0)} \int \frac{\mathbf{j}(\mathbf{r}_1)}{|\mathbf{r}-\mathbf{r}_1|} d\mathbf{r}_1. \qquad (10)$$

Relationship between the current density and total vector potential in the ring is given by the London equation:

$$\mathbf{j} = \frac{\Phi_0 n}{2\pi\mu_0 \lambda_{ab}^2 \rho} \mathbf{i}_\varphi - \frac{1}{\mu_0 \lambda_{ab}^2}(\mathbf{A}_j + \mathbf{A}_L), \qquad (11)$$

where $\Phi_0 = \pi\hbar/e$ is the fluxoid, $n$ is the number of fluxoids trapped into the ring.

Unlike the vector-potential $\mathbf{A}_j$ generated by the Cooper pairs at their regular circular motion in the supercurrent states, the vector-potential $\mathbf{A}_L(\mathbf{r})$ is created by the intrinsic magnetic moments of the Cooper pairs. The vector potential at the point $\mathbf{r}$ generated by the magnetic moment of the $i$–Cooper pair, is:

$$\hat{\mathbf{A}}_i(\mathbf{r}) = \frac{\mu_0}{4\pi} \frac{[\hat{\boldsymbol{\mu}}_i, \mathbf{r}-\mathbf{r}_i]}{|\mathbf{r}-\mathbf{r}_i|^3}, \qquad (12)$$

where $\hat{\boldsymbol{\mu}}_i$ is the operator of the magnetic moment of the pair. The total vector potential is $\mathbf{A}_L = \sum_i \mathbf{A}_i(\mathbf{r})$, where the summation is taken over all the Cooper pairs. Using (2), (3) and (12), we find the average value of the total vector potential:

$$\mathbf{A}_L = A_L(\rho)\mathbf{i}_\varphi = \frac{1}{4\pi}\frac{T_0}{T}\int \frac{[\mathbf{B}(\mathbf{r}_1), \mathbf{r}-\mathbf{r}_1]}{|\mathbf{r}-\mathbf{r}_1|^3} d\mathbf{r}_1, \qquad (13)$$

where the integration is carried out over the ring volume.

Thus, we have the closed system of two equations. Considering (10), (11) and (13), the first equation is:

$$\mathbf{j}(\mathbf{r},T) = \frac{\Phi_0 n}{2\pi\mu_0 \lambda_{ab}^2 \rho}\mathbf{i}_\varphi - \frac{T}{4\pi\lambda_{ab}^2(T-T_0)}\int\frac{\mathbf{j}(\mathbf{r}_1)}{|\mathbf{r}-\mathbf{r}_1|}d\mathbf{r}_1 - \frac{1}{4\pi\mu_0\lambda_{ab}^2}\frac{T_0}{T}\int\frac{[\mathbf{B}(\mathbf{r}_1),\mathbf{r}-\mathbf{r}_1]}{|\mathbf{r}-\mathbf{r}_1|^3}d\mathbf{r}_1, \qquad (14)$$

and the second one is given by:

$$\mathbf{B}(\mathbf{r},T) = \frac{\mu_0 T}{4\pi(T-T_0)}\int\frac{[\mathbf{j}(\mathbf{r}_1),\mathbf{r}-\mathbf{r}_1]}{|\mathbf{r}-\mathbf{r}_1|^3}d\mathbf{r}_1. \qquad (15)$$

Substituting (15) into (14), the latter reduces to the Fredholm integral equation of the second kind for the current density in the superconducting film. In turn, (15) becomes the equation for the magnetic induction in the hole and film regions.

## 4. Observables

We analyze the low temperature behaviours of three quantities, which, in principle, can experimentally be measured. They are the heat capacity and magnetic moment of the ring, and magnetic induction in the ring center.

Except for the lattice and electron energy of superconductor, the ring energy is composed of the magnetic energy, the kinetic energy of the supercurrent, the Zeeman energy (2) and the energy of the magnetic dipole interaction of Cooper pairs (3):

$$E_r(T) = \frac{1}{2}\int d\mathbf{r}\left(\mathbf{BH} + \mu_0\lambda_{ab}^2\mathbf{j}^2\right) + W_Z + W_d. \tag{16}$$

From (11) we have

$$2\pi\rho\left(\mu_0\lambda_{ab}^2 j(\rho,T) + A_j(\rho,T) + A_L(\rho,T)\right) = \Phi_0 n \tag{17}$$

for any $a \leq \rho \leq b$. Using (17), the kinetic and magnetic energies in (16) are reduced to the form:

$$\frac{1}{2}\int d\mathbf{r}\left(\mathbf{BH} + \mu_0\lambda_{ab}^2\mathbf{j}^2\right) = \frac{1}{2}\Phi_0 nJ - \frac{1}{2}\int \mathbf{j}\mathbf{A}_L d\mathbf{r}, \tag{18}$$

where $J$ is the total current, $J = d\int j d\rho$. Using (13) and then (10), the last term in the right-hand of (18) is rewritten as:

$$-\frac{1}{2}\int \mathbf{j}\mathbf{A}_L d\mathbf{r} = -\frac{\pi d}{\mu_0}\frac{T_0(T-T_0)}{T^2}\int_a^b \rho B^2(\rho)d\rho. \tag{19}$$

The Zeeman energy (3) can be presented as:

$$W_Z = -2\pi d\int_a^b n_C \mu_z B(\rho)\rho d\rho. \tag{20}$$

Using (5) and (8), the integral in (20) is reduced to the form:

$$W_Z = -\frac{2\pi d}{\mu_0}\frac{T_0}{T}\int_a^b \rho B^2(\rho)d\rho. \tag{21}$$

Now we estimate the energy (4). Taking into account (5), we obtain:

$$W_d = 2\mu_0\left(\frac{\mu_C^2}{kT}\right)^2 \sum_{i\neq j}\frac{B(\mathbf{r}_i)B(\mathbf{r}_j)}{|\mathbf{r}_i-\mathbf{r}_j|^3}\left[1 - 3\frac{(z_i-z_j)^2}{|\mathbf{r}_i-\mathbf{r}_j|^2}\right]. \tag{22}$$

The interaction (4) or (22) falls off rapidly with distance between pairs $|\mathbf{r}_i - \mathbf{r}_j|$. Therefore, we can restrict the interaction between neighboring magnetic moments. The distance between them is $|\mathbf{r}_i - \mathbf{r}_j| \approx n_C^{-1/3}$. Of course, the magnetic field in the ring is almost constantly at such distances. As a result we have:

$$W_d \approx \frac{\pi d}{\mu_0}\left(\frac{T_0}{T}\right)^2 \int_a^b \rho B(\rho)^2 d\rho \qquad (23)$$

Comparing (21) with (23), we conclude that $W_d / E_L \approx T_0/T \ll 1$. Thus, taking into account of (18) - (21) and (23), the ring energy (17) at $T \gg T_0$ can finally be written as:

$$E_r = \frac{1}{2}\Phi_0 n d \int_a^b j(\rho,T) d\rho - 3\frac{\pi d}{\mu_0}\left[\frac{T_0}{T-T_0} + O\left(\left(\frac{T_0}{T}\right)^2\right)\right]\int_a^b \rho B^2 d\rho. \qquad (24)$$

Using (24), we find the heat capacity of the ring

$$C_r(T) = \frac{\partial E_r}{\partial T}. \qquad (25)$$

Note that in addition to (25), the total heat capacity of the cuprate ring includes the specific heat of the electron and phonon systems that has the dependence $\gamma_{el}T + \gamma_{ph}T^3$ at low temperatures [17]. As shown in Section 6, the ring capacity (25) has a very different temperature dependence, and, hence, can be easily separated from the electron and phonon contributions.

Using (15) and taking into account that the thickness of the ring is much smaller than its inner radius ($a \gg d$), in the hole center the magnetic induction is given by:

$$B(\rho = 0, T) = \frac{\mu_0 d}{2}\frac{T}{T-T_0}\int_a^b j(\rho,T)\frac{d\rho}{\rho}. \qquad (26)$$

Besides the characteristics mentioned above, the magnetic moment of the ring can also be measured experimentally [18]. Taking into account (6), it given by:

$$M_r(T) = \pi d \int_a^b \rho^2 j d\rho + \frac{2\pi d}{\mu_0}\frac{T_0}{T}\int_a^b \rho B d\rho. \qquad (27)$$

## 5. Details of calculations

For flat thin-film rings with thickness $d < \lambda_{ab}$ the current density is usually assumed to be independent of the variable $z$. However, in equations (14) and (15) the variables $z$ and $z_1$ cannot be omitted, because it would lead to the divergence of the current and field at the inner and outer boundary of the ring. To avoid this, we used the following approach. In the equations of the closed system (14)-(15), we performed the integration over $z_1$ for three values of $z = -d/2, 0, d/2$, and the average of the three results was taken.

We introduce the dimensionless variables:

$$\tilde{j}(\tilde{\rho} = \rho/\lambda_{ab}) = j(\tilde{\rho})/j_0, \quad \tilde{B}(\tilde{\rho}) = B/B_0, \quad \tilde{a} = a/\lambda_{ab}, \tilde{b} = b/\lambda_{ab}, \tilde{d} = d/\lambda_{ab}, \qquad (28)$$

where

$$j_0 = \frac{\Phi_0 n}{\mu_0 \lambda_{ab}^3}, \quad B_0 = \frac{\Phi_0 n}{\lambda_{ab}^2}.$$

In these variables the equation (14) is rewritten as:

$$\tilde{j}(\tilde{\rho}) = \frac{1}{2\pi\tilde{\rho}} - \frac{T}{(T-T_0)} \int_{\tilde{a}}^{\tilde{b}} F(\tilde{\rho}, \tilde{\rho}_1) \tilde{j}(\tilde{\rho}_1) d\tilde{\rho}_1 - \frac{T_0}{T} \int_{\tilde{a}}^{\tilde{b}} G(\tilde{\rho}, \tilde{\rho}_1) \tilde{B}(\tilde{\rho}_1) d\tilde{\rho}_1, \qquad (29)$$

where

$$F(\tilde{\rho}, \tilde{\rho}_1) = \frac{\tilde{\rho}_1}{6\pi} \int_0^\pi \cos\varphi_1 d\varphi_1 \ln \frac{\left(\tilde{d}/2 + \sqrt{\tilde{d}^2/4 + a^2}\right)\left(\tilde{d} + \sqrt{\tilde{d}^2 + a^2}\right)}{\left(-\tilde{d}/2 + \sqrt{\tilde{d}^2/4 + a^2}\right)\left(-\tilde{d} + \sqrt{\tilde{d}^2 + a^2}\right)}, \qquad (30)$$

$$G(\tilde{\rho}, \tilde{\rho}_1) = \frac{\tilde{d}\tilde{\rho}_1}{6\pi} \int_0^\pi d\varphi_1 \frac{\tilde{\rho} - \tilde{\rho}_1 \cos\varphi_1}{\tilde{\rho}^2 + \tilde{\rho}_1^2 - 2\tilde{\rho}\tilde{\rho}_1 \cos\varphi_1} \left\{ \frac{1}{\left(\tilde{d}^2/4 + a^2\right)^{1/2}} + \frac{2}{\left(\tilde{d}^2 + a^2\right)^{1/2}} \right\} \qquad (31)$$

with the function $a^2 = \tilde{\rho}^2 + \tilde{\rho}_1^2 - 2\tilde{\rho}\tilde{\rho}_1 \cos\varphi_1$.

Respectively, the equation (15) is reduced to:

$$\tilde{B}(\tilde{\rho}) = \frac{T}{T-T_0} \int_{\tilde{a}}^{\tilde{b}} H(\tilde{\rho}, \tilde{\rho}_1) \tilde{j}(\tilde{\rho}_1) d\tilde{\rho}_1, \qquad (32)$$

where

$$H(\tilde{\rho}, \tilde{\rho}_1) = \frac{\tilde{d}\tilde{\rho}_1}{6\pi} \int_0^\pi d\varphi_1 \frac{\tilde{\rho}_1 - \tilde{\rho} \cos\varphi_1}{\tilde{\rho}^2 + \tilde{\rho}_1^2 - 2\tilde{\rho}\tilde{\rho}_1 \cos\varphi_1} \left\{ \frac{1}{\left(\tilde{d}^2/4 + a^2\right)^{1/2}} + \frac{2}{\left(\tilde{d}^2 + a^2\right)^{1/2}} \right\}. \qquad (33)$$

From (29) and (32) we obtain the integral equation for the current density in the ring:

$$\tilde{j}(\tilde{\rho}) = \frac{1}{2\pi\tilde{\rho}} - \int_{\tilde{a}}^{\tilde{b}} S(\tilde{\rho}, \tilde{\rho}_1) \tilde{j}(\tilde{\rho}_1) d\tilde{\rho}_1 \qquad (34)$$

with the kernel

$$S(\tilde{\rho}, \tilde{\rho}_1) = \frac{T}{T-T_0} F(\tilde{\rho}, \tilde{\rho}_1) + \frac{T_0}{T-T_0} \int_{\tilde{a}}^{\tilde{b}} G(\tilde{\rho}, \tilde{\rho}_2) H(\tilde{\rho}_2, \tilde{\rho}_1) d\tilde{\rho}_2. \qquad (35)$$

Passing from the variable $\tilde{\rho}$ to its $N$ discrete values $\tilde{\rho}_i$, the kernel (35) is transformed to the $N \times N$ matrix $S_{ki}$, and the equation (34) reduces to a system of $N$ linear equations for discrete values of the current $\tilde{j}_i$. After solving this system, from the equation (32) we found the discrete values of the magnetic field $\tilde{B}_i$ and, then, the characteristics (25), (26) and (27).

All calculations are made for the ring parameters: $a = 10\lambda_{ab}$, $b = 20\lambda_{ab}$, and the ring thickness $d = \lambda_{ab}/2$. For results presented in Figs.1 and 2 of the next section, the relative temperature in units of $T_0$ is used. To calculate the heat capacity, the characteristic temperature (8) is requested. For results shown in Fig.3, it is assumed that $T_0 = 2\,mK$ which is consistent with the estimation in Section 3.

## 6. Results and their discussions

Fig. 1 presents the current density and magnetic induction distributions for two temperatures. The temperature dependence of the current is surprising because the current increases with temperature. This behavior can be understood from the London equation (17). The vector potential created by the magnetic moments of the Cooper pairs ($\mathbf{A}_L(\mathbf{r})$ in the left-hand side of (17)) decreases with increasing temperature. Since the right-side of (17) is a constant magnetic flux trapped in the ring, the current must increase with temperature.

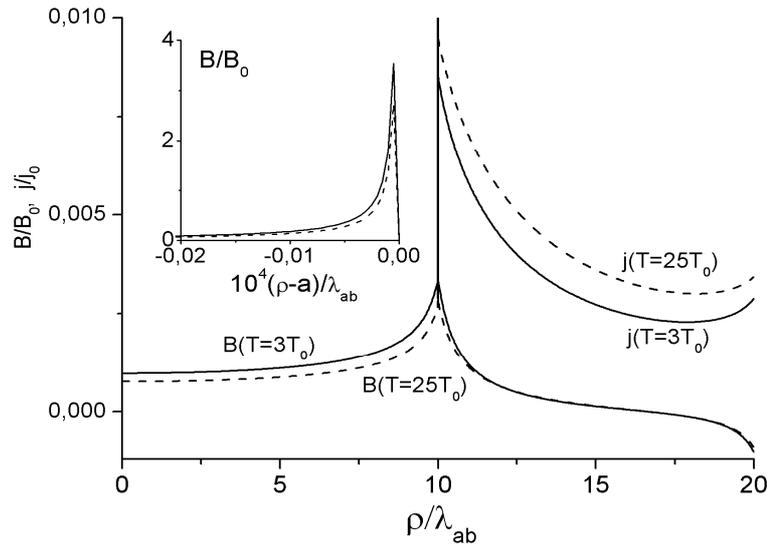

Fig. 1. Profiles of the current $j$ in the ring and magnetic induction $B$ for two temperatures. Inset: the magnetic induction near the inner edge of the ring. The solid curves correspond to $T = 3T_0$, and dashed curves - $T = 25T_0$.

The magnetic induction near the center of the ring is almost constant. Upon approaching to the inner edge of the ring the field increases sharply, as shown in the inset in Fig. 1. An important point is that the magnetic induction reaches the maximum value slightly left of the inner edge of the ring. Inside the ring the magnetic induction is much lower compared to this maximum.

In contrast to the current behavior, the magnetic induction in the ring hole decreases with increasing temperature. This feature follows from the equation (32). For the results shown in Fig. 1, the factor before the integral in (32) decreases by 1.44 times with increasing temperature from $3T_0$ to $25T_0$, but the corresponding increase of the current only partially compensates this temperature factor.

As noted in Section 2, the magnetic field into the ring must be less than the lower critical field $B_{c1}$ for the superconductor. For the $3T_0$ data in Fig.1 the maximum field is $\widetilde{B}(\widetilde{a}) = 3.29*10^{-3}$. Taking into account (28), we obtain the limit on the number of fluxoids trapped into the ring,

$$n < n_c = \frac{B_{c1}\lambda_{ab}^2}{\widetilde{B}(\widetilde{a})\Phi_0}.$$

For estimation we take $B_{c1} = 20\,\text{mT}$ which was experimentally obtained for the $YBa_2Cu_3O_{7-\delta}$ thin film at $T = 0.8T_c$ [19], and $\lambda_{ab} = 2*10^3\,\text{Å}$ [20]. Then we have $n_c = 118$.

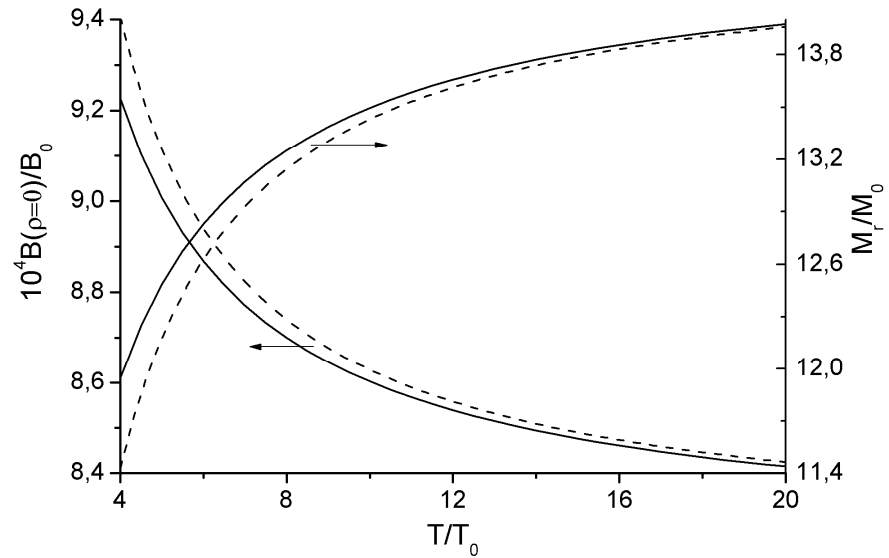

Fig. 2. Temperature dependences of the magnetic induction in the ring center and magnetic moment of the ring. The solid curves are calculated from the equations (26) and (27), dashed curves are the approximations (41) and (43).

Fig. 2 demonstrates the temperature dependences of the magnetic induction (26) in the ring center and the ring magnetic moment (27). This field decreases and the moment increases monotonically with increasing temperature. Both these quantities tend to the asymptotic values that correspond to the total destruction of the ordering of the magnetic moments of Cooper pairs (the solid lines in Fig. 2). This asymptotic behavior is determined by the characteristic temperature $T_0$ (8). The problem is how to extract $T_0$ from these temperature dependences.

For this purpose, at $T \gg T_0$ we seek the approximate solution of the integral equation (34) in the form

$$\tilde{j}(\tilde{\rho}) = \tilde{j}^{(0)}(\tilde{\rho}) + \frac{T_0}{T - T_0} \tilde{j}^{(1)}(\tilde{\rho}), \tag{36}$$

where $\tilde{j}^{(0)}(\tilde{\rho})$ and $\tilde{j}^{(1)}(\tilde{\rho})$ do not depend on temperature.

Substituting (36) in (34) and taking into account of (35), we obtain:

$$\tilde{j}^{(0)}(\tilde{\rho}) + \int_{\tilde{a}}^{\tilde{b}} F(\tilde{\rho}, \tilde{\rho}_1) \tilde{j}^{(0)}(\tilde{\rho}_1) d\tilde{\rho}_1 = \frac{1}{2\pi\tilde{\rho}} \tag{37}$$

and

$$\tilde{j}^{(1)}(\tilde{\rho}) + \int_{\tilde{a}}^{\tilde{b}} F(\tilde{\rho}, \tilde{\rho}_1) \tilde{j}^{(1)}(\tilde{\rho}_1) d\tilde{\rho}_1 = -\int_{\tilde{a}}^{\tilde{b}} D(\tilde{\rho}, \tilde{\rho}_1) \tilde{j}^{(0)}(\tilde{\rho}_1) d\tilde{\rho}_1, \tag{38}$$

where

$$D(\tilde{\rho}, \tilde{\rho}_1) = F(\tilde{\rho}, \tilde{\rho}_1) + \int_{\tilde{a}}^{\tilde{b}} G(\tilde{\rho}, \tilde{\rho}_2) H(\tilde{\rho}_2, \tilde{\rho}_1) d\tilde{\rho}_2. \tag{39}$$

Using (37)-(40), we can find the expression for the magnetic field in the ring center. Considering that the thickness of the ring is much smaller than its inner radius ($a \gg d$), the function (33) is reduced to:

$$\tilde{H}(\tilde{\rho} = 0, \tilde{\rho}_1) = \frac{d}{2\tilde{\rho}_1}. \tag{40}$$

Substituting (40) in (32) and using (36), we have:

$$\tilde{B}(0, T) = \tilde{B}^{(0)} + \tilde{\xi}_B \frac{T_0}{T - T_0}, \tag{41}$$

with

$$\tilde{B}^{(0)} = \frac{d}{2} \int_{\tilde{a}}^{\tilde{b}} \tilde{j}^{(0)}(\tilde{\rho}) \frac{d\tilde{\rho}}{\tilde{\rho}}, \quad \tilde{\xi}_B = \frac{d}{2} \int_{\tilde{a}}^{\tilde{b}} \left[\tilde{j}^{(0)}(\tilde{\rho}) + \tilde{j}^{(1)}(\tilde{\rho})\right] \frac{d\tilde{\rho}}{\tilde{\rho}}. \tag{42}$$

With account of (36) the magnetic moment of the ring (27) in units of $M_0 = \mu_0^{-1}\Phi_0 n \lambda_{ab}$ is

$$\tilde{M}_r(T) = \tilde{M}_r^{(0)} + \tilde{\xi}_m \frac{T_0}{T - T_0}, \qquad (43)$$

where

$$\tilde{M}_r^{(0)} = \pi \tilde{d} \int_{\tilde{a}}^{\tilde{b}} \tilde{\rho}^2 \tilde{j}^{(0)}(\tilde{\rho}) d\tilde{\rho}, \quad \tilde{\xi}_m = \pi \tilde{d} \int_{\tilde{a}}^{\tilde{b}} \tilde{\rho}\left(\tilde{\rho}\tilde{j}^{(1)} + 2\tilde{B}^{(0)}\right) d\tilde{\rho}. \qquad (44)$$

Here $\tilde{B}^{(0)}(\rho)$ is given by (32) with replacements of $T/(T-T_0) \to 1$ and $\tilde{j} \to \tilde{j}^{(0)}$.

The parameters (42) and (44) do not depend on temperature, and are only determined by the geometrical sizes of the ring in units of the penetration depth. For the ring considered, our calculations yielded $\tilde{B}^{(0)} = 8.24*10^{-4}$ and $\tilde{\xi}_B = 3.50*10^{-4}$ in equation (42), and $\tilde{M}^{(0)} = 14.43$ and $\tilde{\xi}_m = -8.94$ in equation (44).

These dependences (41) and (43) are presented by the dash curves in Fig. 2. At $T \leq 10T_0$ the approximation (36) is clearly not sufficient, and the correction of the order of $(T_0/(T-T_0))^2 \tilde{j}^{(2)}$ should be taken into account. However, in this case one must also consider the magnetic dipole interaction of the Cooper pairs (4), as discussed in Section 4. At $T > 12T_0$ both the dash curves demonstrate reasonable agreement with the corresponding solid curves.

Thus, to extract the characteristic temperature $T_0$, the observables $B(\rho = 0, T)$ and $M_r(T)$ can be approximated by the dependences (41) and (43) in the temperature range $T > 12T_0$. A better agreement can be achieved for the derivatives of these values with respect to temperature.

The temperature dependence of the ring energy (24) in units of $E_0 = \Phi_0^2 n^2 / \mu_0 \lambda_{ab}$ is presented in Fig. 3. The correction $O\left((T_0/T)^2\right)$ in the right side of (24) was omitted. This energy increases with temperature, and tends to an asymptotic value. Such behavior is determined by the characteristic temperature (8).

It is convenient to pass to the molar heat capacity. From (24) and (25) it can be written as:

$$C_{r\mu} = \frac{\mu}{\rho V_r} \frac{\partial E_r}{\partial T}, \qquad (45)$$

where $\mu$ and $\rho$ are the molar weight and density of the superconductor, $V_r$ is the ring volume. Also, using (24) and (36), we obtain the approximation of (45) that can be applied to determine $T_0$,:

$$C_{r\mu}^a = \frac{\mu}{\rho V_r} \frac{\Phi_0^2 n^2}{\mu_0 \lambda_{ab}} \frac{\tilde{\xi}_E T_0}{(T-T_0)^2}, \qquad (46)$$

where

$$\tilde{\xi}_E = \frac{\tilde{d}}{2} \int_{\tilde{a}}^{\tilde{b}} \left( 6\pi\tilde{\rho}\tilde{B}^{(0)2} - \tilde{j}^{(1)} \right) d\tilde{\rho}. \qquad (47)$$

The parameter (47) does not depend on temperature, and determined only by the geometrical sizes of the ring in units of the penetration depth. For the ring considered, our calculations yielded $\tilde{\xi}_E = 6.80*10^{-3}$.

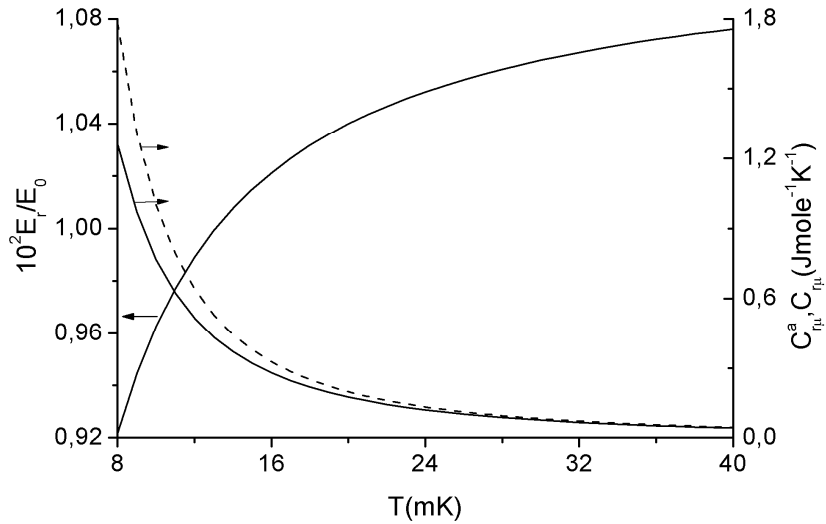

Fig. 3. The temperature dependence of the ring energy and molar heat capacity. The superconductor parameters: the penetration depth $\lambda_{ab} = 2*10^3 \overset{0}{A}$, the molar weight $\mu = 665 g/mole$, the density $\rho = 6.3 g/cm^3$, and the characteristic temperature $T_0 = 2mK$. The ring parameters: $a=2\mu$, $b=4\mu$, $d=10^3 \overset{0}{A}$. The fluxoid number is $n=100$. The solid curves are calculated from the equations (24) and (25), the dashed curve is the approximation of the specific heat (46)-(47).

The temperature dependences of the heat capacities (45) and (46) calculated for the superconducting ring made of $YBa_2Cu_3O_{7-\delta}$, are also shown in Fig. 3. Noteworthy the specific heat is large compared to the electron and phonon contributions. The calculated dependence (45) is in good agreement with the approximation (46) at $T > 10T_0$.

## 7. Conclusion

The pairing symmetry in superconductors determines the intrinsic magnetic moment of the Cooper pairs. It is shown that the magnetic moments of the Cooper pairs can be found from the low-temperature dependences of the characteristics of the thin-film rings made of the cuprate superconductors. These measurable quantities, which were calculated in this work, are the magnetic induction in the ring center, the heat capacity and magnetic moment of the superconducting ring.

If for a given superconductor the predicted low-temperature dependencies of these quantities are not observed experimentally, then the Cooper pairs have no magnetic moment and, accordingly, the characteristic temperature $T_0$ vanishes. Therefore, the pairing wave function has isotropic *s*-wave symmetry. In the opposite case, the Cooper pairs have a non-zero magnetic moment. The characteristic temperature is given by the magnetic moment of Cooper pairs and their concentration. The latter can be determined independently.

Thus, the experimental determination of the magnetic moment of the Cooper pairs can shed light on the pairing symmetry in the cuprates. Unlike the known phase-sensitive experimental techniques, which study the symmetry of the order parameter in the high-$T_c$ superconductor, the proposed method allows to determine the pairing symmetry of the Cooper pairs. In general, these two symmetries may be different.